\documentclass[11pt,twoside]{article}
\usepackage{cozumel2005}
\usepackage{epsf}
\usepackage{psfig}
\usepackage{lscape}
\def\ni{\noindent}                                       
\def\etal{et\thinspace al.\ }                               

\newcommand{\Ha}{\ifmmode {\rm H}\alpha \else H$\alpha$\fi\xspace}
\newcommand{\Hb}{\ifmmode {\rm H}\beta \else H$\beta$\fi\xspace}
\newcommand{\Hg}{\ifmmode {\rm H}\gamma \else H$\gamma$\fi\xspace}
\newcommand{\Hd}{\ifmmode {\rm H}\delta \else H$\delta$\fi\xspace}

\newcommand{\Hii}{\ifmmode \rm{H}\,\textsc{ii} \else H~{\sc ii}\fi}

\newcommand{\nii}{\ifmmode [\rm{N}\,\textsc{ii}] \else [N~{\sc ii}]\fi}

\newcommand{\oi}{\ifmmode [\rm{O}\,\textsc{i}] \else [O~{\sc i}]\fi}

\newcommand{\neiii}{\ifmmode [\rm{Ne}\,\textsc{iii}] \else [Ne~{\sc iii}]\fi}
\newcommand{\hei}{\ifmmode [\rm{He}\,\textsc{i}] \else [He~{\sc i}]\fi}
\newcommand{\oii}{\ifmmode [\rm{O}\,\textsc{ii}] \else [O~{\sc ii}]\fi}

\newcommand{\oiii}{\ifmmode [\rm{O}\,\textsc{iii}] \else [O~{\sc iii}]\fi}

\newcommand{\sii}{\ifmmode [\rm{S}\,\textsc{ii}] \else [S~{\sc ii}]\fi}
\newcommand{\siii}{\ifmmode [\rm{S}\,\textsc{iii}] \else [S~{\sc iii}]\fi}

\begin{document}
\title{Spectral Synthesis of SDSS Galaxies}    
\author{L. Sodr\'e Jr.$^1$, R. Cid Fernandes$^2$, A. Mateus$^1$, G. 
Stasi\'nska$^3$, J. M. Gomes$^2$}   
\affil{$^1$Instituto de Astronomia, Geof\'{\i}sica e Ci\^encias
         Atmosf\'ericas, Universidade de S\~ao Paulo, S\~ao Paulo, SP,
         Brazil \\
$^2$Depto.\ de F\'{\i}sica - CFM - Universidade Federal de
Santa Catarina, SC, Brazil \\
$^3$LUTH, Observatoire de Meudon, 92195 Meudon Cedex, France}    

\begin{abstract} 
We investigate the power of spectral synthesis
as a mean to estimate physical properties of galaxies. Spectral
synthesis is nothing more than the decomposition of an observed
spectrum in terms of a superposition of a base of simple stellar
populations of various ages and metallicities (here from Bruzual \& Charlot 
2003), producing as output the
star-formation and chemical histories of a galaxy, its extinction and
velocity dispersion. We discuss the reliability of this approach 
and apply it to a volume limited sample of 50362
galaxies from the SDSS Data Release 2, producing a catalog of stellar
population properties. Emission lines are also studied, their
measurement being performed after subtracting the computed starlight
spectrum from the observed one. A comparison with recent estimates of
both observed and physical properties of these galaxies obtained by
other groups shows good qualitative and quantitative agreement,
despite substantial differences in the method of analysis. The
confidence in the method is further strengthened by several empirical
and astrophysically reasonable correlations between synthesis results
and independent quantities.  For instance, we report the existence of
strong correlations between stellar and nebular metallicites, stellar
and nebular extinctions, mean stellar age and equivalent width of
H$\alpha$ and 4000 \AA\ break, and between stellar mass and velocity
dispersion. We also present preliminary results of an analysis of a
magnitude-limited sample  which
clearly reveals that the bimodality of galaxy populations is present in
the parameters computed in the synthesis. Our results are also consistent
with the ``down-sizing'' scenario of galaxy formation and evolution.
Finally, we point out one of the major problems facing spectral synthesis
of early-type systems: the spectral base adopted here is based on 
solar-scaled evolutionary tracks whose abundance pattern
may not be appropriate for this type of galaxy.
\end{abstract}

\section{Introduction}

\label{sec:Introduction}

In the proceedings of the meeting on Stellar Populations held in Baltimore
in 1986 Searle (1986), discussing  integral light spectral synthesis, states
that ``this subject has a bad reputation. Too much has been claimed,
and too few have been persuaded''. 
Indeed, recovering the stellar content of a galaxy from its observed
integrated spectrum is not an easy task, as can be deduced from the
amount of work devoted to this topic over the past half
century.  The situation is however much more favorable nowadays. Huge
observational and theoretical efforts in the past few years have
produced large sets of high quality spectra of stars (e.g.,
Prugniel \& Soubiran 2001; Le Borgne \etal 2003; Bertone \etal 2004;
Gonz\'alez-Delgado \etal 2005). These libraries are being implemented
in a new generation of evolutionary synthesis models, allowing the
prediction of galaxy spectra with an unprecedented level of detail
(Vazdekis 1999; Bruzual \& Charlot 2003, hereafter BC03; 
Le Borgne \etal 2004).  At the same time, galaxy
spectra are now more abundant than ever. 
The Sloan Digital Sky Survey (SDSS), in particular, is
providing a homogeneous data base of hundreds of thousands of
galaxy spectra in the 3800--9200 \AA\ range, with a resolution of
$\lambda / \Delta \lambda \sim 1800$ (York \etal 2000; Stoughton \etal
2002; Abazajian \etal 2003, 2004).  This enormous amount of high
quality data will undoubtedly be at the heart of tremendous progress
in our understanding of galaxy constitution, formation and
evolution. Actually, galaxy spectra encode information on the age 
and metallicity
distributions of the constituent stars, which in turn reflect the
star-formation and chemical histories of the galaxies. Retrieving this
information from observational data in a reliable way is crucial for a
deeper understanding of galaxy formation and evolution and, in fact,
significant steps in this direction have recently
been made (Kauffmann et al. 2003a, hereafter K03; 
Brinchmann \etal 2004; Tremonti \etal 2004; Heavens
\etal 2004; Panter, Heavens \& Jimenez 2004).

Here we demonstrate that, besides providing excellent
starlight templates to aid emission line studies, spectral synthesis
recovers reliable stellar population properties out of galaxy spectra
of realistic quality. We show that this simple method provides robust
information on the stellar age ($t_\star$) and stellar metallicities
($Z_\star$) distributions, as well as on the extinction, velocity
dispersion and stellar mass. The ability to recover information on
$Z_\star$ is particularly welcome, given that stellar metallicities
are notoriously more difficult to assess than other properties.  In
order to reach this goal we follow: (1) {\it a priori} arguments,
based on simulations; (2) comparisons with independent work based on a
different method, and (3) an {\it a posteriori} empirical analysis of
the consistency of results obtained for a large sample of SDSS
galaxies. Most of the results presented here are discussed in detail
in Cid Fernandes et al. (2005).

In Section 2 we
present an overview of our synthesis method and simulations designed
to test it and evaluate the uncertainties involved. The discussion is
focused on how to use the synthesis to derive robust estimators of
physically interesting stellar population properties. Section
3 defines a volume limited sample of SDSS
galaxies and presents the results of the synthesis of their spectra,
along with measurements of emission lines.  We also compare in this section
our  results to those obtained by other authors. 
Stellar population and emission
line properties are used in Section 4 to
investigate whether the synthesis produces astrophysically plausible
results. In Section 5 we present preliminary results of an analysis
of a magnitude-limited sample of 20000 galaxies also extracted from SDSS.
Finally, Section 6 summarizes our main
findings.

\section{Spectral Synthesis}

\label{sec:Synthesis}

\subsection{Method}

\label{sec:SynthesisMethod}

Our synthesis code, which we call STARLIGHT, was first discussed in
Cid Fernandes et al. (2004, hereafter CF04). 
We fit an observed spectrum $O_\lambda$ with a
combination of $N_\star$ Simple Stellar Populations (SSP) from the
evolutionary synthesis models of BC03.  Extinction is modeled as due
to foreground dust, and parametrized by the V-band extinction
$A_V$. The Galactic extinction law of Cardelli, Clayton \& Mathis
(1989) with $R_V =3.1$ is adopted. Line of sight stellar motions are
modeled by a Gaussian distribution $G$ centered at velocity $v_\star$
and with dispersion $\sigma_\star$.  With these assumptions the model
spectrum is given by

\begin{equation}
\label{eq:model}
M_\lambda = M_{\lambda_0}
    \left[
    \sum_{j=1}^{N_\star} x_j b_{j,\lambda} r_\lambda
    \right]
    \otimes G(v_\star,\sigma_\star)
\end{equation}

\ni where $b_{j,\lambda}$ is the spectrum of the $j^{\rm th}$ SSP
normalized at $\lambda_0$, $r_\lambda \equiv 10^{-0.4 (A_\lambda -
A_{\lambda_0})}$ is the reddening term, $M_{\lambda_0}$ is the
synthetic flux at the normalization wavelength, $\vec{x}$ is the {\it
population vector} and $\otimes$ denotes the convolution operator.
Each component $x_j$ ($j = 1\ldots N_\star$) represents the fractional
contribution of the SSP with age $t_j$ and metallicity\footnote{In
this paper we follow the convention used in stellar evolution studies,
which define stellar metallicities in terms of the fraction of mass in
metals. In this system the Sun has $Z_\star = 0.02$.} $Z_j$ to the
model flux at $\lambda_0$. The base components can be equivalently
expressed as a mass fractions vector $\vec{\mu}$.  In this work we
adopt a base with $N_\star = 45$ SSPs, encompassing 15 ages between
$10^6$ and $1.3 \times 10^{10}$ yr and 3 metallicities: $Z = 0.2$, 1
and 2.5 $Z_\odot$. Their spectra  were
computed with the STELIB library (Le Borgne \etal 2003), Padova 1994
tracks, and Chabrier (2003) IMF (see BC03 for details).

The fit is carried out with a simulated annealing plus Metropolis
scheme which searches for the minimum $\chi^2 = \sum_\lambda \left[
\left(O_\lambda - M_\lambda \right) w_\lambda \right]^2$, where
$w_\lambda^{-1}$ is the error in $O_\lambda$.  Regions around emission
lines, bad pixels or sky residuals are masked out by setting
$w_\lambda = 0$. Pixels which deviate by more than 3 times the rms
between $O_\lambda$ and an initial estimate of $M_\lambda$ are also
given zero weight.

Fig. 1
illustrates the spectral fit obtained for a galaxy drawn from the
SDSS database. The top-left panel shows the observed spectrum (thin
line) and the model (thick), as well as the error spectrum (dashed).
The bottom-left panel shows the $O_\lambda - M_\lambda$ residual
spectrum, while the panels in the right summarize the derived
star-formation history encoded in the age-binned population vector.
This example, along with those in K03
and CF04, demonstrates that
this simple method is capable of reproducing real galaxy spectra to an
excellent degree of accuracy.

An important application of the synthesis is to measure emission
lines from the residual spectrum, as done by K03. Another, of course, 
is to infer stellar
population properties from the fit parameters.  Our central goal here is to  
investigate
whether spectral synthesis can also recover reliable stellar
population properties. In the remainder of this section we address
this issue by means of simulations.

\begin{figure}[!ht]
\plotone{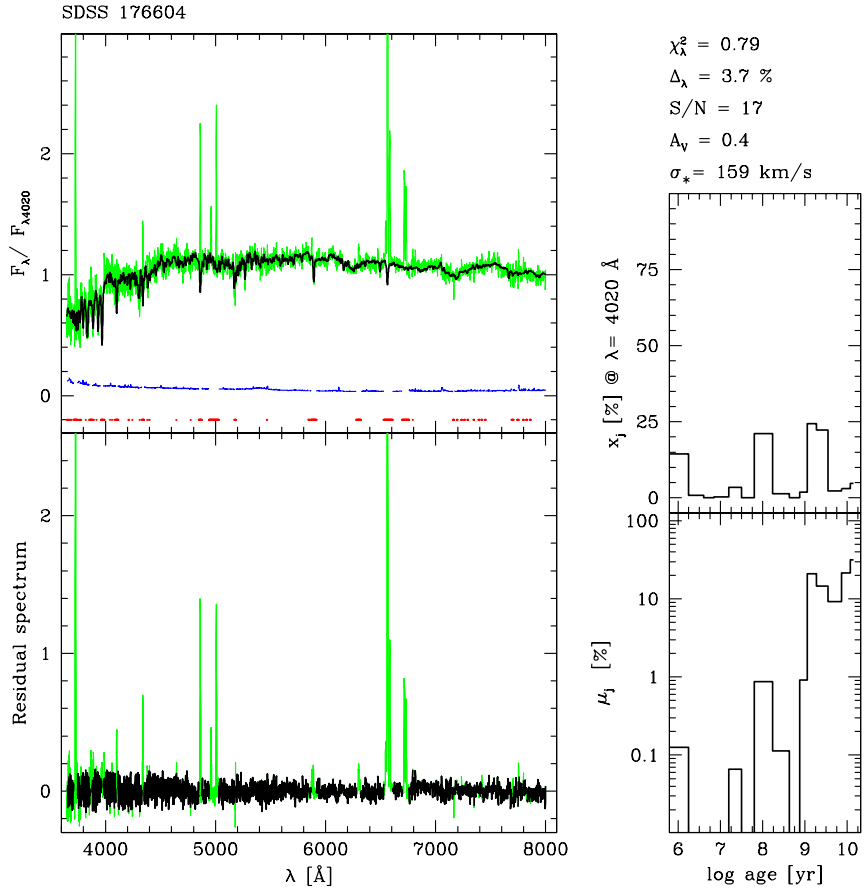}
\caption{Spectral synthesis of a late type SDSS galaxy. Top-left:
Observed (thin line), model (thick line) and error spectra
(dashed line). Points along a horizontal line in the bottom of this
panel indicate 
bad pixels (as given by
the SDSS flag) or emission lines windows, both of which were masked
out in the fits. Bottom-left: Residual spectrum. Masked regions are
plotted with a lighter thin line. Right: Flux (top) and mass (bottom)
fractions as a function of age.  Some of the derived properties are
listed in the top-right.  $\chi^2_\lambda$ is the reduced $\chi^2$,
and $\Delta_\lambda$ is the mean relative difference between model and
observed spectra; $S/N$ refers to the region around $\lambda_0 = 4020$
\AA.}
\label{fig:examples_fits_2}
\end{figure}

\subsection{Robust description of the synthesis results}

\label{sec:RobustDecription}

The existence of multiple solutions is an old known problem in stellar
population synthesis, and even superb
spectral fits as that shown in Fig. 1
do not guarantee that the resulting
parameter estimates are trustworthy.
There is, consequently, a need to assess the degree to
which one can trust the parameters involved in the fit before using
them to infer stellar populations properties. The
spectral fits involve $N_\star + 3$ parameters: $N_\star - 1$ of the
$\vec{x}$ components (one degree of freedom is removed by the
normalization constraint), $M_{\lambda_0}$, $A_V$ and the two
kinematical parameters, $v_\star$ and $\sigma_\star$. The reliability
of parameter estimation is best studied by means of simulations which
feed the code with spectra generated with known parameters, add noise,
and then examine the correspondence between input and output values.

\subsubsection{Simulations}

\label{sec:Simulations}

We have carried out simulations designed to test the method and
investigate which combinations of the parameters provide robust
results. Several sets of simulations were performed. Given our interest in
modeling SDSS galaxies, here we focus on simulations tailored to match
the characteristics of this data set. Test galaxies were built from
the average $\vec{x}$, $A_V$ and $\sigma_\star$ within 65 boxes in the
mean stellar age versus mean stellar metallicity plane obtained for
the sample described in Section 3. 
These new simulations confirm previous results reported in CF04 
that the individual components of
$\vec{x}$ are very uncertain, so we skip a detailed comparison between
$\vec{x}_{\rm input}$ and $\vec{x}_{\rm output}$ and jump straight to
results based on more robust descriptions of the synthesis output.

\subsubsection{Condensed population vector}

A coarse but robust description of the star-for\-ma\-ti\-on history of a
galaxy may be obtained by binning $\vec{x}$ onto ``Young'' ($t_j <
10^8$ yr), ``Intermediate-age'' ($10^8 \le t_j \le 10^9$ yr), and
``Old'' ($t_j > 10^9$ yr) components ($x_Y$, $x_I$ and $x_O$
respectively).  These age-ranges were defined on the basis of the
simulations, by seeking which combinations of $x_j$'s produce smaller
input $-$ output residuals.  
Our simulations show that these 3 components are
very well recovered by the method, with uncertainties smaller than
$\Delta x_Y = 0.05$, $\Delta x_I = 0.1$, and $\Delta x_O = 0.1$ for
$S/N \ge 10$.

\subsubsection{Mass, extinction and velocity dispersion}

\label{sec:Mass_and_vd}

We have also analyzed the input versus output
values of $A_V$, $\sigma_\star$ and the stellar mass $M_\star$. The
latter is not an explicit input parameter of the models, but may be
computed from $\vec{\mu}$ and the $M_\star/L_{\lambda_0}$ ratio of the
different populations in the base. The uncertainties in the recovery of these
parameters are $\Delta A_V < 0.05$ mag, $\Delta \log M_\star < 0.1$
dex and $\Delta \sigma_\star < 12$ km$\,$s$^{-1}$ for $S/N \ge 10$.

\subsubsection{Mean stellar age}

\label{sec:Mean_Stellar_Age}

If one had to choose a single parameter to characterize the stellar
population mixture of a galaxy, the option would certainly be its mean
age. We define two versions of mean stellar age (the logarithm of the age,
actually), one weighted by light

\begin{equation}
\label{eq:mean_age_flux}
<\log t_\star>_L =  \sum_{j = 1}^{N_\star} x_j \log t_j
\end{equation}

\ni and another weighted by stellar mass

\begin{equation}
\label{eq:mean_age_mass}
<\log t_\star>_M =  \sum_{j = 1}^{N_\star} \mu_j \log t_j
\end{equation}

\ni Note that, by construction, both definitions are limited to the 1
Myr--13 Gyr range spanned by the base. The mass weighted mean age is
in principle more physical, but, because of the non-constant $M/L$ of
stars, it has a much less direct relation with the observed spectrum
than $<\log t_\star>_L$.

The simulations show that the mean age is a very robust quantity. The rms
difference between input and output $<\log t_\star>_L$ values is $\le
0.08$ dex for $S/N > 10$, and $\le 0.14$ dex for $<\log t_\star>_M$.
Although the uncertainties
of $<\log t_\star>_L$ and $<\log t_\star>_M$ are comparable in
absolute terms, the latter index spans a smaller dynamical range
(because of the large $M/L$ ratio of old populations), so in practice
$<\log t_\star>_L$ is the more useful of the two indices.

\subsubsection{Mean stellar metallicity}

\label{sec:Mean_Stellar_Z}

Given an option of what to choose as a second parameter to describe a
mixed stellar population, the choice would likely be its typical
metallicity.  Analogously to what we did for ages, we define both
light and mass-weighted mean stellar metallicities:

\begin{equation}
\label{eq:mean_Z_flux}
<Z_\star>_L =  \sum_{j = 1}^{N_\star} x_j Z_j
\end{equation}

\ni and

\begin{equation}
\label{eq:mean_Z_mass}
<Z_\star>_M =  \sum_{j = 1}^{N_\star} \mu_j Z_j
\end{equation}

\ni both of which are bounded by the 0.2--2.5 $Z_\odot$ base limits.
Our simulations  show that the rms of $\Delta \log
<Z_\star>_M = \log <Z_\star>_{M,\rm output} - \log <Z_\star>_{M,\rm
input}$ is of order 0.1 dex. In absolute terms this is comparable to
$\Delta <\log t_\star>$, but note that $<Z_\star>$ covers a much
narrower dynamical range than $<\log t_\star>$, so that in practice
mean stellar metallicities are more sensitive to errors than mean
ages. This is not surprising, given that age is the main driver of
variance among SSP spectra, metallicity having a ``second-order''
effect (e.g., Schmidt \etal 1991; Ronen, Aragon-Salamanca \& Lahav
1999). This is the reason why studies of the stellar populations of
galaxies have a much harder time estimating metallicities than ages,
to the point that one is often forced to bin-over the $Z$ information
and deal only with age-related estimates such as $<\log t_\star>$ (e.g.,
Cid Fernandes \etal 2001; Cid Fernandes, Le\~ao \& Rodrigues Lacerda
2003; K03).

Notwithstanding these notes, it is clear that uncertainties of $\sim
0.1$ dex in $<Z_\star>$ are actually good news, since they do allow us
to recover useful information on an important but hard to measure
property. This new tracer of stellar metallicity is best applicable to
large samples of galaxies such as the SDSS. The statistics of samples
help reducing uncertainties associated with $<Z_\star>$ estimates for
single objects and allows one to investigate correlations between
$<Z_\star>$ and other galaxy properties (Sodr\'e \etal, in
preparation; Section \ref{sec:Stellar_X_Nebular_Metallicities}).

\subsubsection{Age-Metallicity degeneracy}

Our method tends to underestimate $<Z_\star>$ for metal-rich systems
and vice-versa. This trend is due to the infamous age-metallicity
degeneracy.  In order to verify to which degree our synthesis is
affected by this well known problem (e.g., Renzini \& Buzzoni 1986;
Worthey 1994; Bressan, Chiosi \& Tantalo 1996) we have examined the
correlation between the output minus input residuals in $<\log
t_\star>_L$ and $\log <Z_\star>_L$. The age-$Z$ degeneracy acts in the
sense of confusing old, metal-poor systems with young, metal-rich ones
and vice-versa, which should produce anti-correlated residuals. 
This anti-correlation is also
present with the mass-weighted mean stellar metallicity $<Z_\star>_M$,
but is not as strong as for $<Z_\star>_L$. On the other hand, the
uncertainty in $<Z_\star>_M$ is always larger than for $<Z_\star>_L$.

The age-$Z$ degeneracy is thus present in our method, introducing
correlated residuals in our $<Z_\star>$ and $<\log t_\star>$ estimates at
the level of up to $\sim 0.1$--0.2 dex. However, none of the results reported
here rely on this level of precision.

\subsubsection{Base limitations}

We have carried out simulations using the $Z = 0.02 Z_\odot$
BC03 SSPs to examine some of the limitations of the base addopted here. 
Galaxies with such low
metallicity are not expected to be present in significant numbers in
the sample described in Section \ref{sec:Sample_definition}, given
that it excludes low luminosity systems like HII galaxies and dwarf
ellipticals (which are also the least metallic ones by virtue of the
mass-metallicity relation).  Still, it is interesting to investigate
what would happen in this case. When synthesized with our 0.2--$2.5
Z_\odot$ base, these extremely metal poor
galaxies are modeled predominantly with the $0.2 Z_\odot$ components,
as intuitively expected.  Moreover, the mismatch in metallicity
introduces non-negligible biases in other properties, like masses,
mean ages and extinction ($M_\star$, for instance, is systematically
underestimated by 0.3 dex). Similar problems should be encountered
when modeling systems with $Z > 2.5 Z_\odot$. These results serve as
a reminder that our base spans a wide but finite range in stellar
metallicity, and that extrapolating these limits has an impact on the
derived physical properties. While there is no straightforward {\it a
priori} diagnostic of which galaxies violate these limits, in general,
one should be suspicious of objects with mean $Z_\star$ too close to
the base limits.

Our simulations demonstrate that we are able of producing reliable
estimates of several parameters of astrophysical interest, at least in
principle.  We must nevertheless emphasize that this conclusion relies
entirely on models and on an admittedly simplistic view of
galaxies. When applying the synthesis to real galaxy spectra, a series
of other effects come into play. For instance, the extinction law
appropriate for each galaxy likely differs from the one used here. 
Similarly, while in the
evolutionary tracks adopted here the metal abundances are scaled from
the solar values, non-solar abundance mixtures are known to occur in
stellar systems (e.g., Trager \etal 2000a,b; see also Section 5), 
not to mention
uncertainties in the SSP models and the always present issue of the
IMF. Accounting for all these effects in a consistent way is not currently
feasible. We mention these caveats not to dismiss simple models, but
to highlight that all parameter uncertainties discussed above are
applicable within the scope of the model.  

Hence, while the
simulations lend confidence to the synthesis method, one might remain
skeptical of its actual power. The next sections further address the
reliability of the synthesis, this time from a more empirical
perspective.

\section{Analysis of a volume-limited galaxy sample}

\label{sec:Syntesis_SDSS}

In this section we apply our synthesis method to a large sample of
SDSS galaxies to estimate their stellar population properties. We also
present measurements of emission line properties, obtained from the
observed minus synthetic spectra. The information provided by the
synthesis of so many galaxies allows one to address a long menu of
astrophysical issues related to galaxy formation and evolution.
Before venturing in the exploration of such issues, however, it is
important to validate the results of the synthesis by as many means as
possible. Hence, the goal of this study is not so much
to explore the physics of galaxies but to provide an empirical test of
our synthesis method.

\subsection{Sample definition}

\label{sec:Sample_definition}

The spectroscopic data used in this work were taken from the
SDSS. This survey provides spectra of objects in a large wavelength
range (3800--9200 \AA) with mean spectral resolution
$\lambda/\Delta\lambda \sim 1800$, taken with 3 arcsec diameter
fibers. The most relevant characteristic of this survey for our study
is the enormous amount of good quality, homogeneously obtained
spectra. The data analyzed here were extracted from the SDSS main
galaxy sample available in the Data Release 2 (DR2; Abazajian \etal
2004). This flux-limited sample consists of galaxies with
reddening-corrected Petrosian $r$-band magnitudes $r \le 17.77$, and
Petrosian $r$-band half-light surface brightnesses $\mu_{50} \le 24.5$
mag arcsec$^{-2}$ (Strauss \etal 2002).

From the main sample, we first selected spectra with a redshift
confidence $\ge 0.35$. Following the conclusions of Zaritsky,
Zabludoff, \& Willick (1995), we have imposed a redshift limit of $z >
0.05$ (trying to avoid aperture effects and biases; see e.g. G\'omez
\etal 2003) and selected a volume limited sample up to $z = 0.1$,
corresponding to a $r$-band absolute magnitude limit of $M(r)= -20.5$.
The absolute magnitudes used here are k-corrected with the help of the
code provided by Blanton \etal (2003; \texttt{kcorrect v3\_2}) and
assuming the following cosmological parameters: $H_0$ = 70 km s$^{-1}$
Mpc$^{-1}$, $\Omega_M = 0.3$ and $\Omega_\Lambda = 0.7$.  We also
restricted our sample to objects for which the observed spectra show a
$S/N$ ratio in $g$, $r$ and $i$ bands greater than 5.  These
restrictions leave us with a volume limited sample containing $50362$
galaxies, which leads to a completeness level of $\sim$ 98.5 per cent.

\subsection{Results of the spectral synthesis}

\label{sec:SDSS_spectral_fits}

All 50362 spectra were brought to the rest-frame (using the redshifts
in the SDSS database), sampled from 3650 to 8000 \AA\ in steps of 1
\AA, corrected for Galactic extinction\footnote{Unlike in the first
data release, the final calibrated spectra from the DR2 are not
corrected for foreground Galactic reddening.} using the maps given by
Schlegel, Finkbeiner \& Davis (1998) and the extinction law of
Cardelli \etal (1989, with $R_V = 3.1)$, and normalized by the median
flux in the 4010--4060 \AA\ region. The $S/N$ ratio in this spectral
window spans the 5--30 range, with median value of 14. Besides the
masks around the lines listed in Section \ref{sec:Simulations}, we exclude
points with SDSS flag $\ge 2$, which signals bad pixels, sky residuals
and other artifacts. After this pre-processing, the spectra are fed
into the STARLIGHT code described in Section \ref{sec:SynthesisMethod}. On
average, the synthesis is performed with $N_\lambda = 3677$ points,
after discounting the ones which are clipped by our $\le 3$ sigma
threshold (typically 40 points) and the masked ones.
The spectral fits are generally very good.

The total stellar masses of the galaxies were obtained from the
stellar masses derived from the spectral synthesis (which correspond
to the light entering the fibers) by dividing them by $(1 - f)$,
where $f$ is the fraction of the total galaxy luminosity in the
$z$-band outside the fiber.  This approach, which neglects stellar
population and extinction gradients, leads to an increase of typically
0.5 in $\log M_\star$.  We did not apply any correction to the
velocity dispersion estimated by the code given that the spectral
resolution of the BC03 models and the data are very similar.

We point out that we did not constrain the extinction $A_V$ to be
positive. There are several reasons for this choice: (a) some objects
may be excessively dereddened by Galactic extinction; (b) some objects
may indeed require bluer SSP spectra than those in the base; (c) the
observed light may contain a scattered component, which would induce a
bluening of the spectra not taken into account by the adopted pure
extinction law; (d) constraining $A_V$ to have only positive values
produces an artificial concentration of solutions at $A_V = 0$, an
unpleasant feature in the $A_V$ distribution. Interestingly, most of
the objects for which we derive negative $A_V$ (typically $-0.1$ to
$-0.3$ mag) are early-type galaxies. These galaxies are dominated by
old populations, and expected to contain little dust. This is
consistent with the result of K03, who find negative extinction
primarily in galaxies with a large $D_n(4000)$. The distribution of
$A_V$ for these objects, which can be selected on the basis of
spectral or morphological properties, is strongly peaked around $A_V =
0$, so that objects with $A_V < 0$ can be considered as consistent
with having zero extinction.  In any case, none of the results
reported in this paper is significantly affected by this choice.

\subsection{Emission line measurements}

\label{sec:EmissionLines}

Besides providing estimates of stellar population properties, the
synthesis models allow the measurement of emission lines from the
``pure-emission'', starlight subtracted spectra $(O_\lambda -
M_\lambda)$. We have measured the lines of
\oii$\lambda\lambda$3726,3729, \oiii$\lambda$4363, H$\beta$,
\oiii$\lambda\lambda$4959,5007, \oi$\lambda$6300, \nii$\lambda$6548,
H$\alpha$, \nii$\lambda$6584 and \sii$\lambda\lambda$6717,6731. Each
line was treated as a Gaussian with three parameters: width, offset
(with respect to the rest-frame central wavelength), and flux.  Lines
from the same ion were assumed to have the same width and offset. We
have further imposed \oiii$\lambda5007$/\oiii$\lambda4959 = 2.97$ and
\nii$\lambda6584$/\nii$\lambda6548 = 3$ flux ratio constraints.
Finally, we consider a line to have significant emission if its fit
presents a $S/N$ ratio greater than 3.

In some of the following analysis, galaxies with emission lines are
classified according to their position in the \oiii/H$\beta$ versus
\nii/H$\alpha$ diagram proposed by Baldwin, Phillips \& Terlevich
(1981) to distinguish normal star-forming galaxies from galaxies
containing active galactic nuclei (AGN). We define as normal
star-forming galaxies those galaxies that appear in this diagram and
are below the curve defined by K03 (see also Brinchmann \etal
2004). Objects above this curve are transition objects and galaxies
containing AGN.

\subsection{Comparisons with the MPA/JHU database}

\label{sec:Comparisons}

The SDSS database has been explored by several groups, using different
approaches and techniques.  The MPA/JHU group has recently publicly
released catalogues\footnote{available at
http://www.mpa-garching.mpg.de/SDSS/} of derived physical properties
for 211894 SDSS galaxies, including 33589 narrow-line AGN (K03, see
also Brinchmann \etal 2004). These catalogues are based on the K03
method to infer the star formation histories, dust attenuation and
stellar masses of galaxies from the simultaneous analysis of the 4000
\AA\ break strength, $D_n(4000)$, and the Balmer line absorption index
H$\delta_A$. These two indices are used to constrain the mean stellar
ages of galaxies and the fractional stellar mass formed in bursts over
the past few Gyr, and a comparison with broad-band photometry then
allows to estimate the extinction and stellar masses.

The MPA/JHU catalogues provide very useful benchmarks for similar
studies. In this section we summarize the  comparison between  
values of some of the
parameters from these catalogues with our own estimates. A more complete
discussion is presented in Cid Fernandes et al. (2005).

\subsubsection{Stellar extinction}

The MPA/JHU group estimates the $z$-band stellar extinction $A_z$
through the difference between model and measured colours, assuming an
attenuation curve proportional to $\lambda^{-0.7}$. In our case, the
extinction $A_V$ is derived directly from the spectral fitting,
carried out with the Milky Way extinction law (Cardelli \etal 1989,
c.f. Section \ref{sec:SynthesisMethod}). 
These two independent estimates are very strongly and linearly
correlated, with a Spearman rank correlation coefficient $r_S =
0.95$. However, the values of $A_z$ reported by the MPA/JHU group
are systematically larger
than our values: $A_z({\rm MPA/JHU})\simeq 2.51 A_z({\rm This~Work})$
in the median. 
This discrepancy, however, is only apparent because 
the Galactic extinction law is
substantially harder than $\lambda^{-0.7}$. One thus expects to need
less extinction when modeling a given galaxy with the former law than
with the latter. Our analysis indicates that there are no
substantial differences between the MPA/JHU and our estimates of the
stellar extinction other than those implied by the differences in the
reddening laws adopted in the two studies.

\subsubsection{Stellar masses}

Our results for the total stellar masses compares very well to the MPA/JHU 
extinction-corrected stellar masses, with $r_S = 0.89$. The quantitative
agreement is also good, with a median difference of just 0.1 dex.
This small offset seems to be due to a
subtle technicality. Whereas we adopt the $M/L$ ratio of the best
$\chi^2$ model, the MPA/JHU group derives $M/L$ comparing the observed
values of the $D_n(4000)$ and $H\delta_A$ indices with a library of
32000 models. Each model is then weighted by its likelihood, and a
probability distribution for $M/L$ is computed. The MPA/JHU mass is
the median of this distribution, which is not necessarily the same as
the best-$\chi^2$ value. 

\subsubsection{Velocity dispersion}

We use the sub-sample of galaxies with active nuclei to compare our
measurements of absorption line broadening due to galaxy velocity
dispersion and/or rotation, $\sigma_\star$, with those of the MPA/JHU
group, since they list this quantity only in their AGN Catalogue.  The Spearman
correlation-coefficient in this case is $r_S=0.91$ and the median of
the difference between the two estimates is just 9 km$\,$s$^{-1}$,
indicating an excellent agreement between both studies.

\subsubsection{Emission lines and nebular metallicities}
\label{sec:Comparison_EmissionLines}

Brinchmann \etal (2004) also provide, in their Emission line
Catalogue, data on emission lines which can be
compared with our own measurements.  We have compared the
fluxes and equivalent widths of H$\alpha$, \nii$\lambda6584$,
\oii$\lambda3727$, H$\beta$ and \oiii$\lambda5007$ as measured by our
code and that obtained by the MPA/JHU group. We do not find any
significant difference between these values; the largest discrepancy
($\sim$ 5 per cent) was found for the equivalent widths of H$\alpha$
and \nii, probably due to different estimates of the continuum level
and the associated underlying stellar absorption. Our emission line
measurements are also in good agreement with those in Stasi\'nska
\etal (2004), who fit the Balmer lines with emission and absorption
components, instead of subtracting a starlight model.

Overall, we conclude that our spectral synthesis method yields
estimates of physical parameters in good agreement with those obtained
by the MPA/JHU group, considering the important differences in
approach and underlying assumptions.

\section{Empirical relations}
\label{sec:Correlations}

Yet another way to assess the validity of physical properties derived
through a spectral synthesis analysis is to investigate whether this
method yields astrophysically reasonable results. In this section we
follow this empirical line of reasoning by comparing some results
obtained from our synthesis of SDSS galaxies (which excludes emission
lines) with those obtained from a direct analysis of the emission
lines.  Our aim here is to demonstrate that our synthesis results do indeed
make sense.

\subsection{Stellar and Nebular Metallicities}
\label{sec:Stellar_X_Nebular_Metallicities}

Our spectral synthesis approach yields estimates of the mean
metallicity of the stars in a galaxy, $<Z_\star>$. The analysis of
emission lines, on the other hand, gives estimates of the present-day
abundances in the warm interstellar medium. Although stellar and
nebular metallicities are not expected to be equal, it is reasonable
to expect that they should roughly scale with each other.

\begin{figure}[!ht]
\plotone{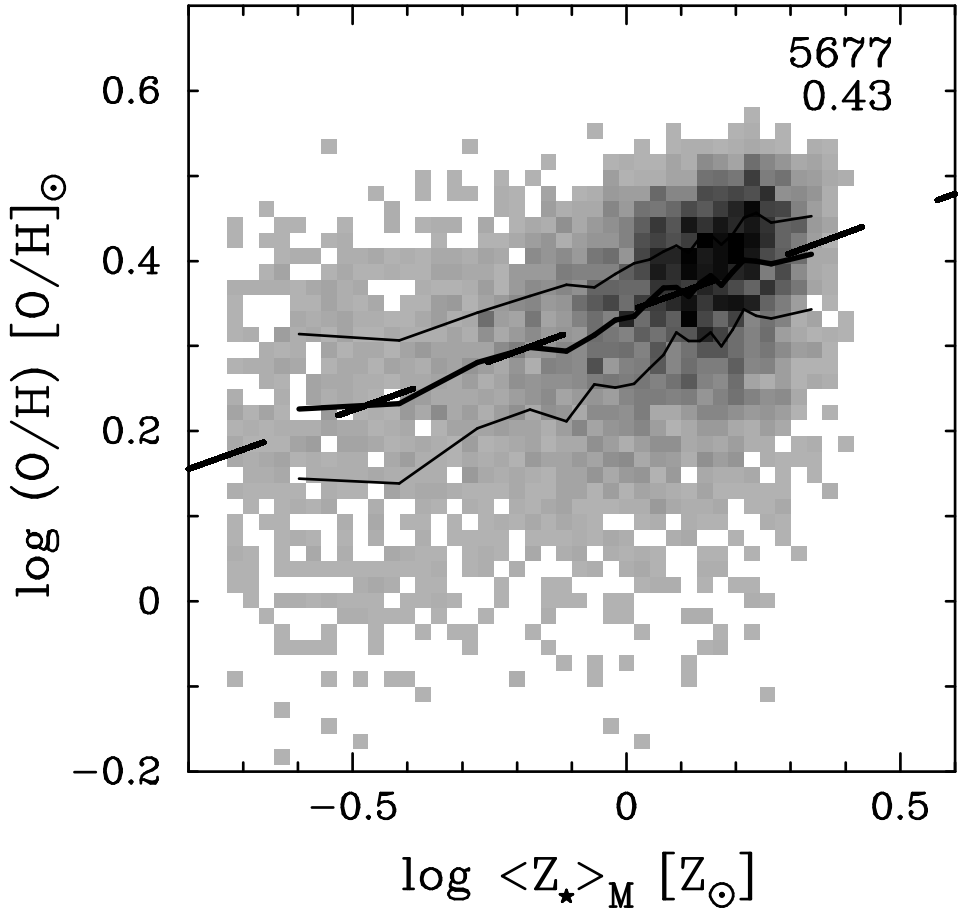}
\caption{Nebular oxygen abundance versus mass-weighted stellar
metallicity, both in solar units, for normal star-forming galaxies in
our sample.  The median values and quartiles in bins of same number of
objects are shown as thin solid lines.  The dashed line is a robust
fit for the relation.}
\label{fig:zneb_zst}
\end{figure}

Fig. \ref{fig:zneb_zst} shows the correlation between mass-weighted
stellar metallicities and the nebular oxygen abundance (computed as in
Section \ref{sec:Comparison_EmissionLines}), both in solar
units\footnote{The solar unit adopted for the nebular oxygen abundance
is $12+\log (\rm{O/H})_\odot=8.69$ (Allende Prieto, Lambert \&
Apslund 2001).}, for our sample of normal star-forming galaxies. A
correlation is clearly seen, although with large scatter ($r_S =
0.42$).  Galaxies with large stellar metallicities also have large
nebular oxygen abundances; galaxies with low stellar abundances tend
to have smaller abundances. The observed scatter is qualitatively
expected due to variations in enrichment histories among galaxies.  

Nebular and stellar metallicities are estimated through completely
different and independent methods, so the correlation depicted in
Fig. \ref{fig:zneb_zst} provides an {\it a posteriori} empirical
validation for the stellar metallicity derived by the spectral
synthesis.  The possibility to estimate stellar metallicities for so
many galaxies is one of the major virtues of spectral synthesis, as it
opens an important window to study the chemical evolution of galaxies
and of the universe as a whole (Panter \etal 2004; Sodr\'e \etal in
prep.).

\subsection{Stellar and Nebular extinctions}

The stellar extinction in the V-band is one of the products of our
STARLIGHT code. A more traditional and completely independent method
to evaluate the extinction consists of comparing the observed
H$\alpha$/H$\beta$ Balmer decrement to the theoretical value, the ``Balmer
extinction'' (e.g., Stasi\'nska \etal 2004).

\begin{figure}[!ht]
\plotone{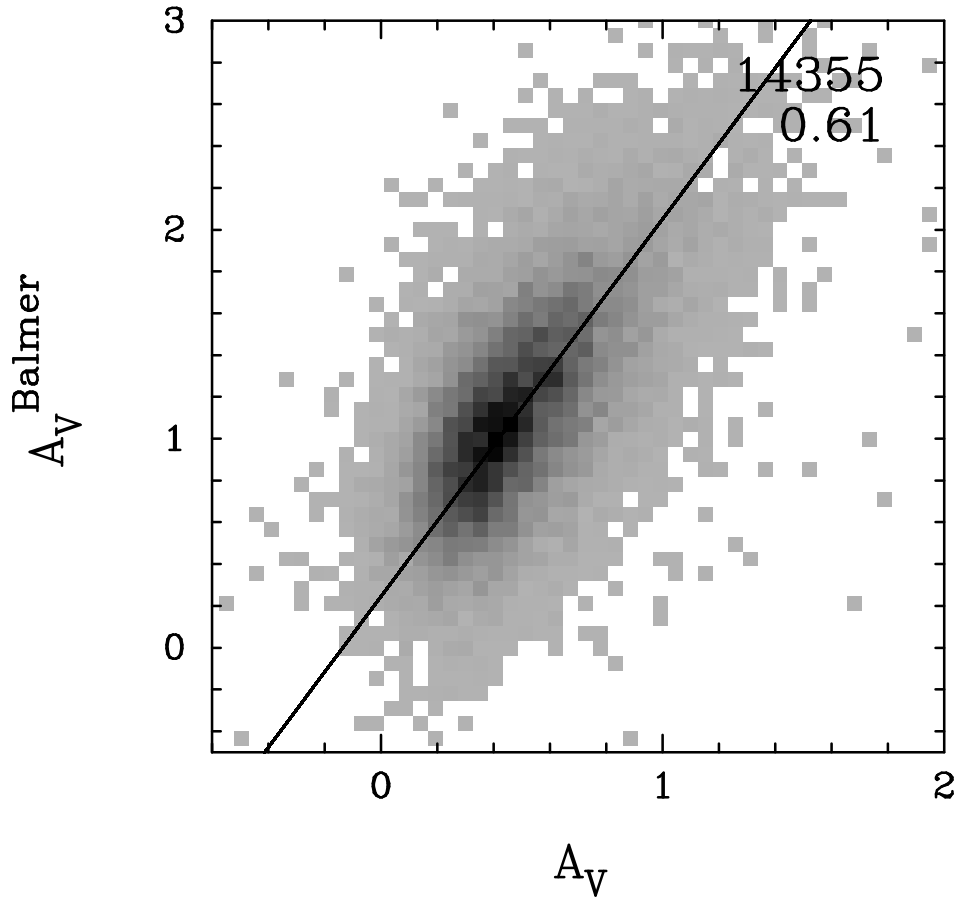}
\caption{Relation between the nebular ($A_V^{\rm{Balmer}}$) and
stellar ($A_V$) extinctions for our sample of normal star-forming galaxies.
The solid line is a robust fit for the relation.}
\label{fig:AV_Chb}
\end{figure}

Fig.~\ref{fig:AV_Chb} presents a comparison between the stellar $A_V$
and $A_V^{\rm Balmer}$.  These two extinctions are determined in
completely independent ways, and yet, our results show that they are
closely linked, with $r_S = 0.61$. A linear bisector fitting yields
$A_V^{\rm Balmer}  = 0.24 + 1.81 A_V$. 
Note that the angular coefficient in this relation indicates that
nebular photons are roughly twice as extincted as the starlight.  This
``differential extinction'' is in very good qualitative and
quantitative agreement with empirical studies (Fanelli \etal 1988;
Calzetti, Kinney \& Storchi-Bergmann 1994; Gordon \etal 1997;
Mas-Hesse \& Kunth 1999). 

\subsection{Relations with mean stellar age}

\begin{figure}[!ht]
\plotone{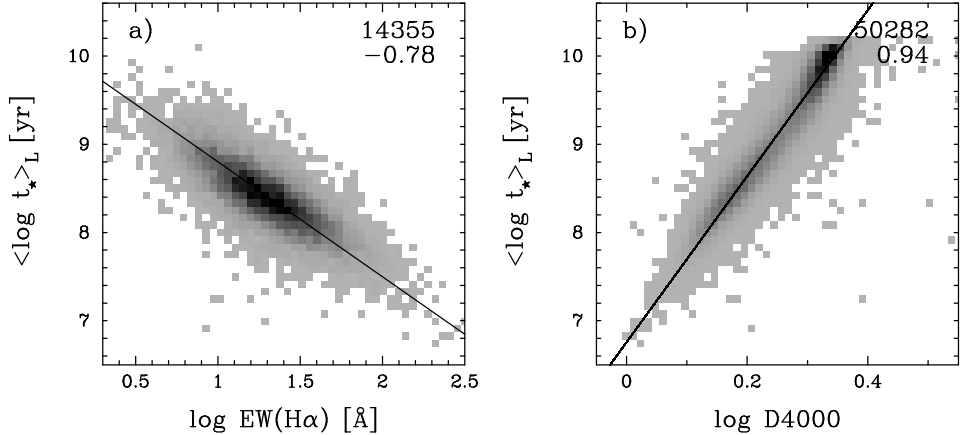}
\caption{(a) Equivalent width of H$\alpha$ versus the light-weighted
mean stellar age for normal emission line galaxies in our sample.  (b)
Relation between the D4000 index and the light-weighted mean
stellar age. The solid lines are robust fits for the relations.}
\label{fig:age_EWHalpha_D4000}
\end{figure}

The equivalent width (EW) of H$\alpha$ is related to the ratio of
present to past star formation rate of a galaxy (e.g., Kennicutt
1998).  It is thus expected to be smaller for older galaxies.
Fig.~\ref{fig:age_EWHalpha_D4000}a shows the relation between
EW(H$\alpha$) and the mean light-weighted stellar age obtained by our
spectral synthesis. The anti-correlation, which has $r_S = -0.78$, is
evident. It is worth
stressing that these two quantities are obtained independently, since
the spectral synthesis does not include emission lines.

Another quantity that is considered a good age indicator, even for
galaxies without emission lines, is the 4000 \AA\ break, D4000.  We
measured this index following Bruzual (1983), who define D4000 as the
ratio between the average value of $F_\nu$ in the 4050--4250 and
3750--3950 \AA\ bands.  The relation between $<\log t_\star>_L$ and
D4000 is shown in Fig.~\ref{fig:age_EWHalpha_D4000}b. Note that the
concentration of points at the high age end reflects the upper age
limit of the base adopted here, 13 Gyr (c.f.\ Section
2.2).  The correlation is very strong ($r_S =
0.94$), showing that indeed D4000 can be used to estimate empirically
mean light-weighted galaxy ages, despite its metallicity dependence
for very old stellar populations (older than 1 Gyr, as shown by
K03). 

\subsection{$M_\star$--$\sigma_\star$ relation}

\label{sec:mass-sigma}

\begin{figure}[!ht]
\plotone{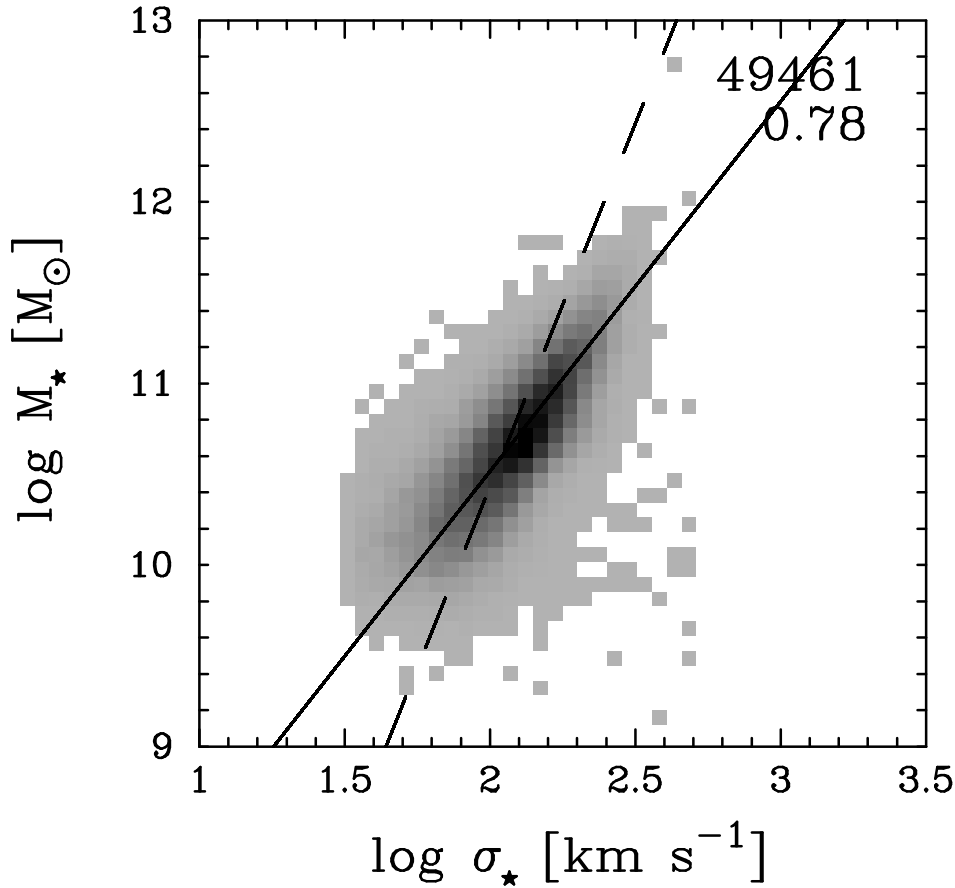}
\caption{Mass-velocity dispersion relation for our sample. The solid
line is a robust fit for the data while dashed one is a fitting
assuming $M_\star \propto \sigma^4$.}
\label{fig:mass-sigma}
\end{figure}

Fig. \ref{fig:mass-sigma} shows the relation between stellar mass and
$\sigma_\star$, obtained from the synthesis.  The relation is quite
good, with $r_S=0.79$. The solid line displayed in the figure is
$\log M_\star = 6.44 + 2.04 \log \sigma_\star$ 
for $M_\star$ in $M_\odot$ and $\sigma_\star$ in km$\,$s$^{-1}$,
obtained with a bisector fitting. The figure also shows as a dashed
line a fit assuming $M_\star \propto \sigma_\star^4$, expected from
the virial theorem under the (unrealistic) assumption of constant mass
surface density. In both cases we have excluded from the fit galaxies
with $\sigma_\star < 35$ km$\,$s$^{-1}$, which corresponds to less
than half the spectral resolution of both data and models.

This is another relation that is expected {\it a priori} if we have in
mind the Faber-Jackson relation for ellipticals and the Tully-Fisher
relation for spirals. For early-type galaxies, $\sigma_\star$ is a
measure of the central velocity dispersion, which is directly linked
to the gravitational potential depth, and, through the virial theorem,
to galactic mass. For late-type systems, $\sigma_\star$ has
contributions of isotropic motions in the bulges, as well as of the
rotation of the disks, and is also expected to relate with galactic
mass. Another aspect that it is interesting to point out in
Fig.~\ref{fig:mass-sigma} is that the dispersion in the
$M_\star$--$\sigma_\star$ relation decreases as we go from
low-luminosity, rotation-dominated systems, for which the values of
$\sigma_\star$ depend on galaxy inclination and bulge-to-disk ratio,
to high-luminosity, mostly early-type systems, which obey a much more
regular (and steeper) relation between $\sigma_\star$ and $M_\star$.

This relation, between a quantity that is not directly linked to the
synthesis, $\sigma_\star$, and another one that is a product of our
synthesis, $M_\star$, is yet another indication that the results of
our STARLIGHT code do make sense.

\section{Recent results}
We have started an analysis of a magnitude-limited sample with 20000
galaxies extracted from DR2, aiming to probe galaxies with luminosities
smaller than those discussed in the previous sections. We present
below a summary of some of the results obtained so far.

\subsubsection{The bimodality of galaxy populations}
With the recent 
advance of redshift surveys, the study of galaxy populations 
has quantitatively revealed  the existence of a bimodal 
distribution in some fundamental galaxy properties, found both in 
photometric and spectroscopic data: star-forming and 
passive galaxies have quite distinct properties. 
Perhaps the most representative bimodal
distribution is that found in galaxy colours. 
Since photometric parameters are easily measured, their bimodal behavior
has been studied  for galaxies in 
the local universe by using both the SDSS and 
the Two Degree Field Galaxy Redshift Survey (2dFGRS) data
(Strateva et al. 2001; Hogg et al. 2002; Blanton et al. 2003, Wild et al. 
2004). A bimodality 
has been found in other galaxy parameters, like mass
(Kauffmann et al. 2003b) and star formation properties
(Madgwick et al. 2002; Wild et al. 2004; Brinchmann et al. 2004). 
We show in Fig. 6 that the bimodal
character of galaxy populations is also clearly present in some parameters
derived from our spectral synthesis (Mateus et al., in prep.). These
results suggest that galaxies have
evolved through two major paths.

\begin{figure}[!ht]
\plotone{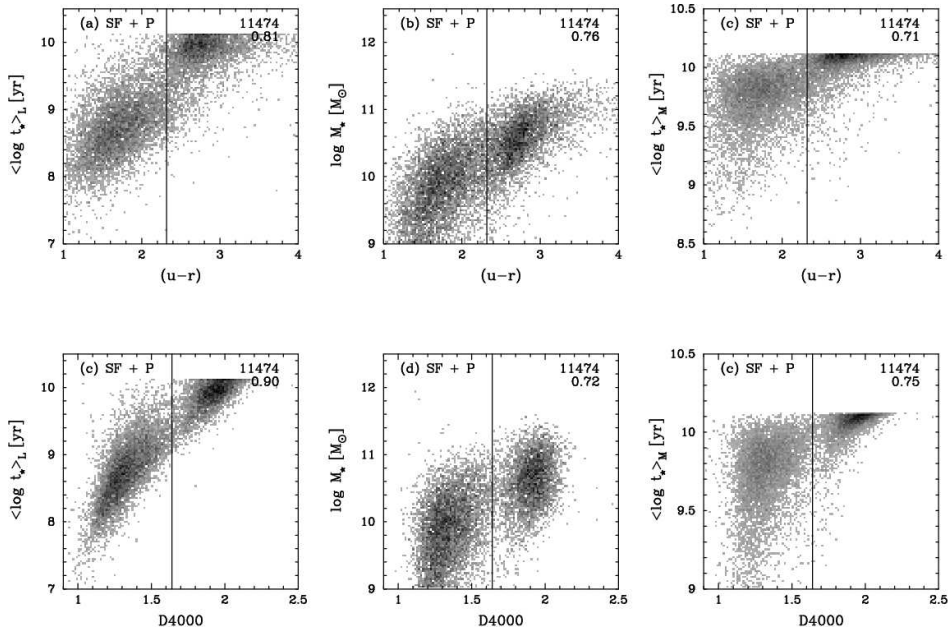}
\caption{Results of the spectral synthesis of 20000 galaxies
of a magnitude-limited
sample extracted from SDSS  showing 
properties of star-forming (SF) and passive (P) galaxies. The vertical lines
correspond to value of the parameters (colours or $D4000$) which provide the
best separation between these two classes of galaxies.}
\end{figure}

\subsubsection{The down-sizing of galaxy populations}
Another interesting result emerging from our synthesis of SDSS galaxies
is the confirmation that there is a correlation between mass and galaxy
age, in the sense that most of the stars in massive galaxies were formed
long time ago, whereas galaxies with a large fraction of young or
intermediate-age stars tend to be less massive. This phenomenon, known
as the down-sizing of galaxy populations (Cowie et al. 1996; Juneau et al.
2004; Kodama et al. 2004), is clearly seem in our results
with this new sample and, actually, is an essential piece of information on
galaxy formation and evolution.

\begin{figure}[!ht]
\plotone{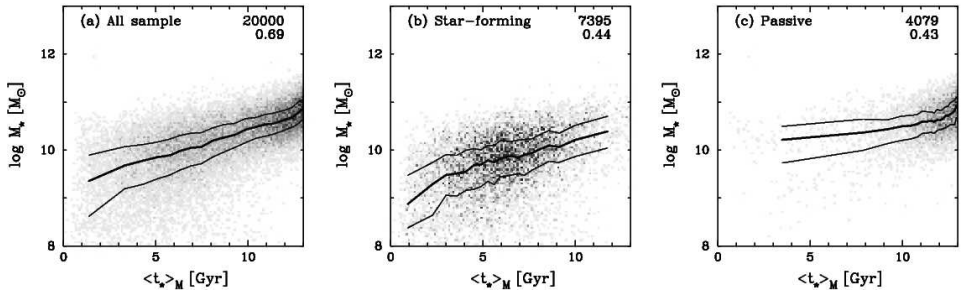}
\caption{Relation between stellar mass and mean mass-weighted stellar ages
for galaxies in the magnitude-limited sample of 20000 galaxies extracted from
SDSS.}
\end{figure}

\subsubsection{The problem of the alpha-enhancement}
A major problem of spectral synthesis with a spectral base using
BC03 spectra is that while their evolutionary tracks are scaled from
the solar values, non-solar abundance patterns may occur in galaxies, mainly
in early-types (e.g., Trager et al. 2000a,b). In practice, as discussed
by BC03, lines associated to $\alpha$-elements are not fitted as well as
other lines. This problem is indeed present in our results, as shown
in Fig. 8 for a subsample of 2488 {\it bona fide} ellipticals with high S/N spectra.
This figure indicates that the residuals in the absorption lines of some
$\alpha$-elements correlate with the depth of the gravitational
potential well of the galaxy (Cid Fernandes et al., in prep.), 
suggesting that our spectral synthesis results
of early-type galaxies may be biased
due to inadequacy of the spectral base adopted here to synthesize this type
of galaxy. 

\begin{figure}[!ht]
\plotone{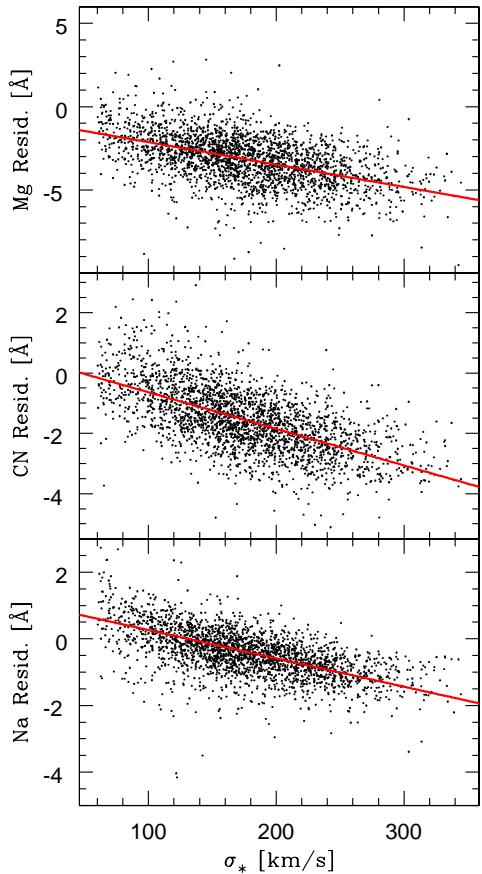}
\caption{Residuals (measured as equivalent widths) of the spectral synthesis
in the region of CN (4132--4196), Mg (5015-5190) and Na D
(5874 -- 5911) lines as a function of the galaxy velocity dispersion. 
The data is for a sample of 2488 early-type galaxies. }
\end{figure}

\section{Summary}

\label{sec:Conclusions}

We have developed and tested a method to fit galaxy spectra with a
combination of spectra of individual simple stellar populations
generated with state-of-the art evolutionary synthesis models. The
main goal of this investigation was to examine the reliability of
physical properties derived in this way. This goal was pursued by
three different means: simulations, comparison with independent
studies, and analysis of empirical results. 

Our simulations show that the individual SSP strengths, encoded in the
population vector $\vec{x}$, are subjected to large uncertainties, but
robust results can be obtained by compressing $\vec{x}$ into coarser
but useful indices. In particular, physically motivated indices such
as mean stellar ages and metallicities are found to be well recovered
by spectral synthesis even for relatively noisy spectra. Stellar
masses, velocity dispersion and extinction are also found to be
accurately retrieved.

We have applied our STARLIGHT code to a volume limited sample of
over 50000 galaxies from the SDSS Data Release 2. The spectral fits
are generally very good, and allow accurate measurements of emission
lines from the starlight subtracted spectrum.  
We have compared our results to those obtained by the MPA/JHU
group (K03; Brinchmann \etal 2004) with a different method to
characterize the stellar populations of SDSS galaxies. The stellar
extinctions and masses derived in these two studies are very strongly
correlated. Furthermore, differences in the values of $A_V$ and
$M_\star$ are found to be mostly due to the differences in the model
ingredients (extinction law).  Our estimates of stellar velocity
dispersions and emission line properties are also in good agreement
with those of the MPA/JHU group.

The confidence in the method is further strengthened by several
empirical correlations between synthesis results and independent
quantities. We find strong correlations between stellar and nebular
metallicites, stellar and nebular extinctions, mean stellar age and
the equivalent width of H$\alpha$, mean stellar age and the
4000 \AA\ break, stellar mass and velocity dispersion. These are all
astrophysically reasonable results, which reinforce the conclusion
that spectral synthesis is capable of producing reliable estimates of
physical properties of galaxies. These results validate spectral 
synthesis as a powerful tool
to study the history of galaxies. 

We have also presented some preliminary results of an analysis of a
magni\-tude-limited sample containing 20000 galaxies from SDSS which
clearly reveal that the bimodality of galaxy populations is present in
the parameters obtained from the synthesis. Our results are also consistent
with the ``down-sizing'' scenario of galaxy formation and evolution.
Finally, we point out one of the major problems of the spectral synthesis
of early-type systems: the spectral base adopted here is based on 
solar-scaled evolutionary tracks which may not be appropriate for this
type of galaxy.

\section*{Acknowledgments}

We congratulate D. Valls-Gabaud and M. Chaves for the organization of
this workshop.  Partial support from CNPq,
FAPESP and the France-Brazil PICS program are also acknowledged.

\end{document}